\documentclass[12pt]{article}
\usepackage[latin1]{inputenc}

\usepackage{amssymb}
\usepackage[dvips]{graphics,color}
\usepackage{epsfig}
\usepackage{float}
\usepackage{bm}

\graphicspath{{figs/}}

\newcommand{\bmu}{\bm\mu}

\begin{document}
%\SIM {1}{4}{00}{28}{00}

%\runningheads{Glimm and L\"{a}uter} {Sample Size Re-estimation}

\begin{center}
 \textbf{\LARGE Some Notes on Blinded Sample Size Re-Estimation}
 %\author{Ekkehard Glimm\affil{1}\comma\corrauth and J\"{u}rgen
%L\"{a}uter\affil{2}}

\vspace{1cm}

Ekkehard Glimm\footnote{Novartis Pharma AG, CH-4002 Basel, Switzerland} and J\"{u}rgen L\"{a}uter\footnote{Otto-von-Guericke-Universit\"{a}t Magdeburg, 39114 Magdeburg, Germany}
\end{center}

\begin{abstract}
This note investigates a number of scenarios in which unadjusted
testing following a blinded sample size re-estimation leads to type
I error violations. For superiority testing, this occurs in certain
small-sample borderline cases. We discuss a number of alternative
approaches that keep the type I error rate. The paper also gives a
reason why the type I error inflation in the superiority context
might have been missed in previous publications and investigates why
it is more marked in case of non-inferiority testing.
\end{abstract}

%\keywords{sample size re-estimation, type I error control, $t$-test}

\section{Introduction}\label{sec1}
Sample Size re-estimation (SSR) in clinical trials has a long
history that dates back to Stein (1945). A sample size review at an
interim analysis aims at correcting assumptions which were made at
the planning stage of the trial, but turn out to be unrealistic.
When the sample units are considered to be normally distributed,
this typically concerns the initial assumption about the variation
of responses. Wittes and Brittain (1990) and Gould and Shih (1992,
1998) among others discussed methods of {\it blinded} SSR. In
contrast to {\it unblinded} SSR, blinded SSR assumes that the
actually realized effect size estimate is not disclosed to the
decision makers who do the SSR. Wittes et al. (1999) and Zucker et
al. (1999) investigated the performance of various blinded and
unblinded SSR methods by simulation. They observed some slight type
I error violations in cases with small sample size and gave
explanations for this phenomenon for some of the unblinded
approaches available at that time.

Slightly later, Kieser and Friede(\cite{Kieser03}, \cite{Friede06})
suggested a method of blinded sample size review which is
particularly easy to implement. In a trial with normally distributed
sample units with the aim of testing for a significant treatment
effect ("superiority testing") at the final analysis, it estimates
the variance under the null hypothesis of no treatment effect and
then proceeds to an unmodified $t$-test in the final analysis, i.e.
a test that ignores the fact that the final sample size was not
fixed from the onset of the trial. Kieser and Friede investigated
the type I error control of their suggestion by simulation. They
conclude that \emph{no additional measures to control the
significance level are required in these designs if the study is
evaluated with the common t-test and the sample size is recalculated
with any of these simple blind variance estimators}.

Although Kieser and Friede explicitly stated that they provide no
formal proof of type I error control, it seems to us that many
statisticians in the pharmaceutical industry are under the
impression that such a proof is available. This, however, is not the
case. In this paper, we show that in certain situations, the method
suggested by Kieser and Friede does not control the type I error.

It should be emphasized that asymptotic type I error control with
blinded SSR is guaranteed. If the sample size of only one of the two
stages tends to infinity, the other stage is obviously irrelevant
for the asymptotic value of the final test statistic and thus the
method asymptotically keeps $\alpha$. If the sample size in both
stages goes to infinity, then the stage-1-estimate of the variance
converges to a constant value. Hence, whatever sample size
re-estimation rule is used, it implicitly fixes the total sample
size in advance (though its precise value is not yet known before
the interim). In any case, asymptotically $\alpha$ is again kept.
Govindarajulu (2003) has formalized this thought and extended to
non-normally distributed data. As a consequence, the type I error
violations discussed in this note are very small and occur in cases
with small samples. We still believe, however, that the statistical
community should be made aware of these limitations of blinded
sample-size review methodology.

While sections \ref{sec2}-\ref{sec4} focus on the common case of
testing for treatment differences in clinical trials, section
\ref{sec5} briefly discusses the case of testing for non-inferiority
of one of the two treatments. In had been noted in another paper by
Friede and Kieser \cite{Friede03} that type I error inflations from
SSR can be more marked in this situation. We give an explanation of
this phenomenon.

\section{A scenario leading to type I error violation}\label{sec2}

In this section we show that in certain cases, a blinded sample size
review as suggested by \cite{Kieser03} leads to a type I error which
is larger than the nominal level $\alpha$.

In general, blinded sample review is characterized by the fact that
the final sample size of the study may be changed at interim
analyses, but that this change depends on the data only via the {\it
total variance} which is the variance estimate under the null
hypothesis of interest. If $x_i, i=1, \ldots, n_1$ are
stochastically independent normally distributed observations, this
total variance is proportional to $\sum_{i=1}^{n_1}x_i^2$ in the
one-sample and to $\sum_{i=1}^{n_1}x_i^2-n_1\bar{x}^2$ in the
two-sample case.

We consider the one-sample $t$ test of $H_0:\mu=0$ at level $\alpha$
applied to $x_i \sim N(\mu, \sigma^2)$. The reason for this is
simplicity of notation and the fact that the geometric
considerations given below cannot be imagined for the two-sample
case which would have to deal with a dimension larger than three
even in the simplest setup. However, the restriction to the
one-sample case entails no loss of generality, as it is conceptually
the same as the two sample case. We will briefly comment on this
further below. In addition, a blinded sample size review may also be
of practical relevance in the one-sample situation, for example in
cross-over trials.

Assume a blinded sample size review after $n_1=2$ observations. If
the total variance is small, we stop sampling and test with the
$n_1=n=2$ observations we have obtained. If it is large, we take
another sample element $x_3$, and do the test with $n=3$
observations. This rule implies that $n=2$ for $x_1^2+x_2^2\leq r^2$
and $n=3$ otherwise for some fixed scalar $r$. Geometrically, the
rejection region of the (one-sided) $t$ test for $n=3$ is a
spherical cone with the equiangular line $x_1=x_2=x_3$ as its
central axis in the three-dimensional space. By definition, the
probability mass of this cone is $\alpha$ under $H_0$. For the case
of $n=2$, the rejection region is a segment of the circle
$x_1^2+x_2^2\leq r^2$ around the equiangular line $x_1=x_2$. Hence,
in three dimensions, the rejection region is a segment of the
spherical cylinder $x_1^2+x_2^2\leq r^2, x_3$ arbitrary. The
probability mass covered by this segment again is $\alpha$ inside
the cylinder. The rejection region of the entire procedure is the
segment of the cylinder plus the spherical cone minus the
intersection of the cone with the cylinder. We now approximate the
probability mass of these components.

For $r^2$ small, we approximately have $P(x_1^2+x_2^2\leq
r^2)=\frac{r^2}{2\sigma^2}$. Hence, under $H_0$, the probability
mass of this part of the rejection region is approximately
$\frac{r^2}{2\sigma^2}\cdot \alpha$. The volume of the intersection
of the cone with the cylinder can be approximated as follows: The
central axis $x_1=x_2=x_3$ of the cone intersects with the cylinder
in one of the points $\pm
\left(\frac{r}{\sqrt{2}},\frac{r}{\sqrt{2}},\frac{r}{\sqrt{2}}\right)$.
The distance of this point to the origin is thus
$h=\sqrt{\frac{3}{2}}r$. The approximate volume of the intersection
is $\frac{4\pi h^3}{3}=\sqrt{6}\pi r^3$. To conservatively
approximate the probability mass of this intersection, we assume
that every point in it has the same probability mass as the origin
(in reality, it of course has a lower probability mass). Then the
probability mass of the intersection is approximated by $\sqrt{6}\pi
r^3\cdot\alpha \cdot (\sqrt{2\pi}\sigma)^{-3}$, where
$(\sqrt{2\pi}\sigma)^{-3}$ is the value of the standard normal
density $N_3\left(\mathbf{0},\sigma^2\mathbf{I}_3\right)$ in the
point $\mathbf{0}$. Combining these results, a conservative
approximation of the probability mass of the rejection region for
the entire procedure is
\begin{equation}\label{f1}
\alpha\left(1+\frac{r^2}{2\sigma^2}-\frac{\sqrt{6}\pi r^3}
{(\sqrt{2\pi}\sigma)^{3}}\right)=\alpha\left(1+\frac{r^2}{2\sigma^2}-\frac{\sqrt{3}
r^3} {2\sqrt{\pi}\sigma^{3}}\right).
\end{equation}
Obviously, this is larger than $\alpha$ for small $r$.

For the more general case of a stage-1-sample size of $n_1$,
possibly followed by a stage 2 with $n_2$ further observations, the
rejection region of the "sample size reviewed" $t$ test has an
approximate null probability following the same basic principle as
(\ref{f1}):\linebreak $\alpha\cdot\left(1+const_1 \cdot
\left(\frac{r}{\sqrt{n_1}\sigma}\right)^{n_1}-const_2 \cdot
\left(\frac{r}{\sqrt{n_1}\sigma}\right)^{n_1+n_2} \right)$ if $r,
n_1$ and $n_2$ are small. Consequently, there must be situations
with small $\frac{r}{\sqrt{n_1}\sigma}$ where the blinded review
procedure cannot keep the type I error level $\alpha$ exactly. Due
to symmetry of the rejection region, this statement holds for both
the one- and the two-sided test of $H_0$.

Note that in this example, the test keeps $\alpha$ exactly if
$\sum_{i=1}^{n_1} x_i^2\leq r^2$. This is due to the sphericity of
the conditional null distribution of $(x_1,\cdots, x_{n_1})$ given
$\sum_{i=1}^{n_1} x_i^2\leq r^2$ (see \cite{Fang90}, theorem 2.5.8).
Type I error violation stems from the fact that the test does not
keep $\alpha$ conditional on $\sum_{i=1}^{n_1} x_i^2> r^2$, i.e. if
a second stage of sampling more observations is done.

To investigate the magnitude of the ensuing type I error violation,
we simulated 10'000'000 cases with $n_1=2$ initial observations and
$n_2=2$ additional observations that are only taken if
$x_1^2+x_2^2\geq 0.5$. The true type I error of the two-sided
combined $t$ test turned out to be $0.0542$ for a nominal
$\alpha=0.05$. As expected, this is caused by the situations where
stage-2-data is obtained. Since $x_1^2+x_2^2 \sim \chi^2(2)$, we
have $P(x_1^2+x_2^2\geq 0.5)=0.779$. This was also the value
observed in the simulations. The rejection rate for these cases
alone was $0.0553$. If $x_1^2+x_2^2< 0.5$, we know that
conditionally the rejection rate is exactly $\alpha$. Accordingly,
this conditional rejection rate in the simulations was $0.0500$.

If $n_1$ and $n_2$ are increased, the true type I error rate
converges rather quickly to $\alpha$. For example, in case of
$n_1=n_2=5$ and $r^2=2.5$, the simulated error rate is $0.0508$ with
$77.6\%$ of cases leading to stage 2 and a conditional error rate of
$0.0510$ in case stage 2 applies.

We also performed some simulations where $n_2$ is determined with
the algorithm suggested by \cite{Kieser03}. For this purpose, we
generated $10'000'000$ simulation runs of a blinded sample size
review after $n_1=2$ observations following the rule given in
section 3 of \cite{Kieser03} with a very large assumed effect of
$\delta=2.2$. This produces an average of $3.09$ additional
observations $n_2$. The simulated type I error was $0.05077$.

To see that the two-sample case is also covered by these
investigations, note that the ordinary $t$-test statistic can be
viewed as $X/\sqrt{Y/s}$ where $X\sim N(\delta,1)$ is stochastically
independent of $Y \sim \chi^2\left(s\right)$. Regarding any
investigation of the properties of this quantity, it obviously does
not matter if the random variables $X$ and $Y$ arise as mean and
variance estimate from a one-sample situation or as difference in
means and common within-group variance estimate in the two-sample
case. The same is true here: According to \cite{Kieser03}, p. 3575,
the "resampled" $t$-test statistic consists of the four components
$D_1$, $V_1$, $D_2\left|\left(V_1,D_1\right)\right.$ and
$V_2^*\left|\left(V_1,D_1\right)\right.$ (loosely speaking, these
correspond to the differences in means and variance estimates of the
two stages). Comparing the distributions of $D_1$ and $V_1$ and the
conditional distributions of $D_2$ and $V_2^*$ given $D_1$ and $V_1$
(and hence $n_2$), one immediately sees that these are the same for
the one- and the balanced two-sample case when we replace $n_i$ by
$n_i/2$ and the means of the two stages by the corresponding two
differences in means between the two treatment groups. For the
conditional distribution of $V_2^*\left|\left(V_1,D_1\right)\right.$
see section \ref{sec4}.

\section{Approaches that control the type I error}

\subsection{Permutation and rotation tests}

If the considerations from the previous section are of concern, then
a simple alternative is to do the test as a permutation test. In the
one-sample case, one would generate all permutations (or a large
number of random permutations) of the signs onto the absolute values
of observations. For each permutation, the $t$ test would be
calculated and the $(1-\alpha)$-quantile of the resulting empirical
distribution of $t$-test values gives the critical value of an exact
level $\alpha$-test of $H_0$. Alternatively, a $p$-value can be
obtained by counting the percentage of values from the permutation
distribution which are larger or equal to the actually observed
value of the test statistic. After determining the additional sample
size $n_2$ from the first $n_1$ observations, we apply the
permutation method to all $n_1+n_2$ observations. The special case
of $n_2=0$ is possible and then the parametric (non-permutation)
$t$-test can also be used. This strategy keeps the $\alpha$-level
exactly, because the total variance
$\frac{1}{n_1}\sum_{i=1}^{n_1}x_i^2$ is invariant to the
permutations.

In the two-sample case, the approach would permute the treatment
allocations of the observations. In order to preserve the observed
total variance,
%this statement is not true. It so happens that this is preserved, but that is no requirement
%for validity of the method:
%as well as the observed ratio between
%number of observations per treatment in stage 1 and stage 2,
the permutations have to be done separately for the $n_1$
observations of stage 1 and the $n_2$ observations of stage 2,
respectively.

%Note: To think about: We need to be able to state that the total variance does not change.
%This is achieved for permutations, but not for rotations (at least not for rotations of all values)!
%Hence, a rotation strategy would have to restrict allowed rotations to those who keep
%$\Sum_{i=1}^{n_1}x_i \geq r^2$. But with this restriction, a rotation strategy would also be ok.
%If $n$ is small, permutations maybe replaced by random rotations of
%the observations (Langsrud, 2005).

If sample sizes are small, permutation tests suffer from the
discreteness of the resampling distribution and the associated loss
of power. In this case, rotation tests \cite{Langs05, Lau05} offer
an attractive alternative. These replace the random permutations of
the sample units by random rotations. This renders the support of
the corresponding empirical distribution continuous and thus avoids
the discreteness problem of the permutation strategy. In order to
facilitate this, rotation tests require the assumption of a
spherical null distribution. This is the case in this context.
Stage-1- and stage-2-data have to be rotated separately even in the
one-sample case in order to keep the fixed observed stage-1-value of
the total variance.

Permutation and rotation strategies emulate the true distribution of
the $t$ test including sample size review. Hence, they will
"automatically" correct any type I error inflation as outlined in
the previous section, but will otherwise have almost identical
properties (e.g. with respect to power) as their "parametric"
counterpart. We did some simulations of the permutation and rotation
strategies under null and non-null scenarios. These, however, just
backed up the statements made here and are thus not reported.

\subsection{Combinations of test statistics from the two stages}

Methods that use a combination of test statistics from the two
stages are another alternative if one is looking for an exact test.
For example, we might use Fisher's $p$-value combination
$-2\log(p_1\cdot p_2)$ \cite{Bauer94} where $p_j=P(T_j>t_j)$ with
$T_j$ being the test statistic from stage-$j$-data only and $t_j$
its observation from the concrete data at hand. As $-2\log(p_1\cdot
p_2) \sim \chi^2(4)$ for independent test statistics $T_1$ and $T_2$
under $H_0$, the combination $p$-value test rejects $H_0$ if
$-2\log(p_1\cdot p_2)$ is larger than the $(1-\alpha)$-quantile from
this distribution. In this application, we use the true null
distributions of the test statistics $T_j$ to determine the
$p$-values. For example, in case of the one-sample-$t$-test these
are the $t$-distributions $T_1 \sim t(n_1-1)$ and $T_2 \sim
t(n_2-1)$.

The stage-2-sample size $n_2$ is uniquely determined by
$\sum_{i=1}^{n_1}x_i^2$.  Since $T_1$ is a test statistic for which
Theorem 2.5.8. of \cite{Fang90} holds under $H_0$, the null
distribution of $T_1$ is valid also conditionally on
$\sum_{i=1}^{n_1}x_i^2$. As a consequence, $T_1 \sim t(n_1-1)$ and
$T_2 \sim t(n_2-1)$ are stochastically independent under $H_0$ for
given $\sum_{i=1}^{n_1}x_i^2$. Any combination of them can be used
as the test statistic for $H_0$. Of course, one still has to find
critical values of the null distribution for the selected
combination.

The statement about the conditional null distributions of the test
statistics given the total variance $\sum_{i=1}^{n_1}x_i^2$ allows
us to go beyond Fisher's $p$-value combination and similar methods
that are combining $p$-values using fixed weights or calculate
conditional error functions with an "intended" stage-2-sample size.
The weights used to combine the two stages may also depend on the
observed stage-1-data. For example, if the variance were known (and
hence a $z$-test for $H_0$ could be done), then the optimal
(standardized) weights for combining the $z$-statistics from the two
stages would be $\sqrt{\frac{n_1}{n_1+n_2}}$ and
$\sqrt{\frac{n_2}{n_1+n_2}}$ in the one-sample case. Hence,
$t_{comb}=\sqrt{\frac{n_1}{n_1+n_2}}t_1+\sqrt{\frac{n_2}{n_1+n_2}}t_2$
seems a promising candidate for a combination test statistic. The
fact that $\left\{T_j\right\}, j=1,2$ retain their $t(n_j-1)$-null
distributions if we condition on $s_1^2=\sum_{i=1}^{n_1}x_i^2$ means
that critical values for this test can be obtained from the
distribution of the weighted sum of two stochastically independent
$t$-distributed random variables with $(n_1-1)$ and $(n_2-1)$
degrees of freedom, respectively. It is obvious that this is very
easy with numerical integration or a simulation. Comparing
$t_{comb}$ with these critical values (that depend only on $n_1$ and
$n_2$) to decide about the rejection of $H_0$ gives an exact
level-$\alpha$ test.

To investigate the performance of the introduced suggestions, we did
several simulations. The critical values for the one-sided
one-sample test using $t_{comb}$ were obtained by simulating
1'000'000 values of two independent $t$-distributions with $n_1$
fixed and $n_2$ as determined by the SSR method in \cite{Kieser03}.
We used the "total variance" for SSR, not the "adjusted variance"
which subtracts a constant based on the putative effect size.
Nevertheless, the re-estimated sample size of course depends on the
"assumed effect" which may be different from the true, unknown
effect size. In the simulations,we investigated various combinations
of the true effect size $\mu$ and an assumed effect size $\delta$.

\begin{figure}[H]
\centering
\begin{tabular}{cc}
 \includegraphics[width=7cm, height=7cm, angle=0]{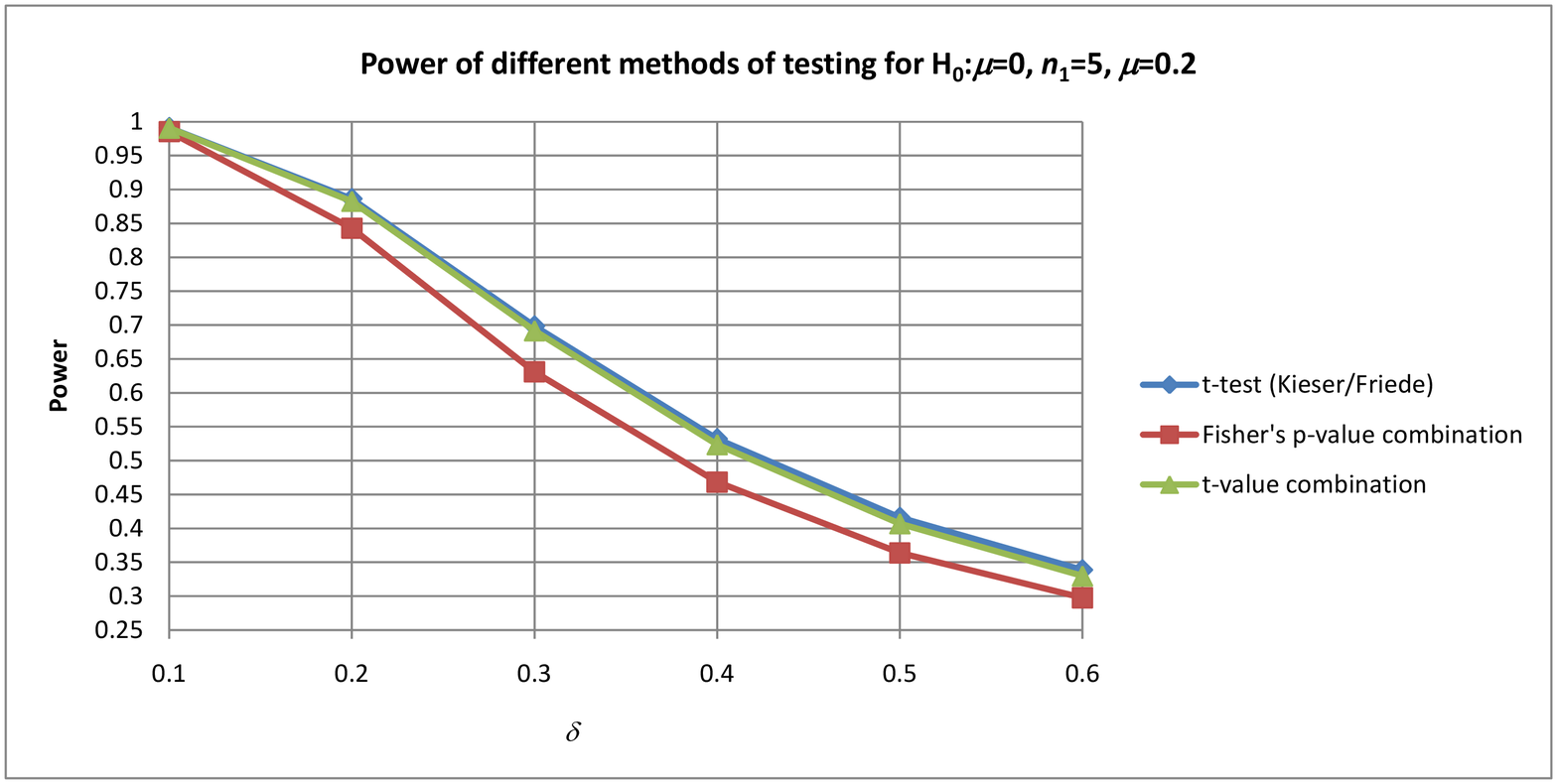} &
 \includegraphics[width=7cm, height=7cm, angle=0]{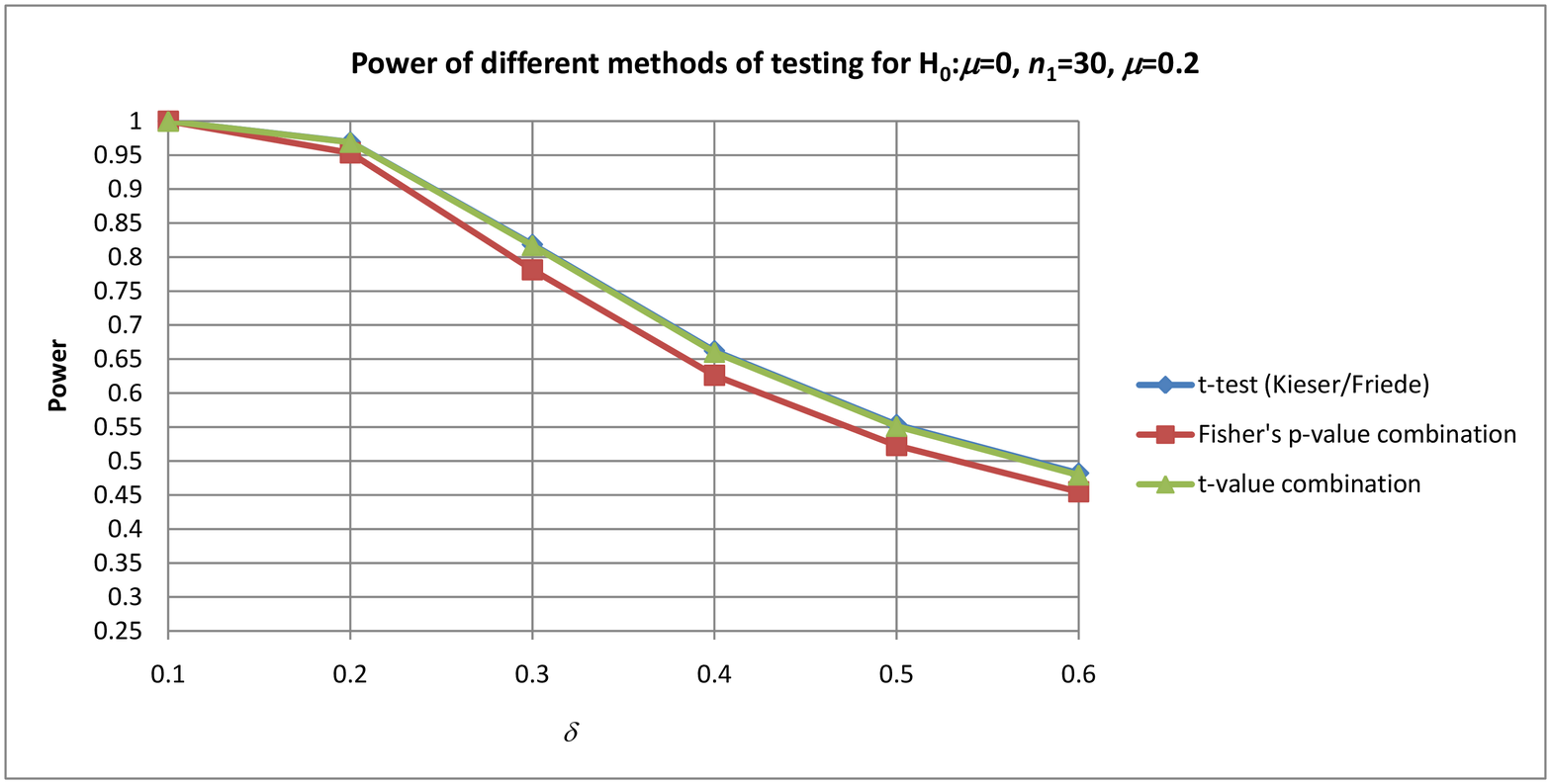}
\end{tabular}
\caption{Power of various test after sample size
re-estimation}\label{fig1}
\end{figure}

Null simulations verified the claimed type-I-error control for the
various adjustment methods described in this section and are thus
not reported. Figure \ref{fig1} shows the results of 1'000'000
simulation runs for sample sizes of $n_1=5$ and $n_1=30$, a true
effect size of $\mu=0.2$ (the standardized true effect size, such
that the non-centrality parameter of a standard-$t$-test with $n$
observations would be $\sqrt{n}\mu$) and varying values of $\delta$
on the $x$-axis. The unmodified $t$-test as suggested by
\cite{Kieser03} is always best. In comparison, the weighted $t$-test
combination $t_{comb}$ suffers from a small power loss which seems
non-negligible only for very small stage-1-sample sizes below
$n_1=10$ (where the type I error control of the "reviewed" $t$-test
might be a concern). For all simulated scenarios with $n_1=30$, the
difference in power was always below $1\%$. In contrast, Fisher's
$p$-value combination typically loses 3 to 4 \% of power when
$t$-test power is less than 95 \% and up to 7\% for some scenarios
($\mu=0.1, \delta=0.15$ with power $t$-test 76.4\%, power
$t$-combination 76.3\%, power $p$-value combination 69.6\%).

%In contrast, if a rotation strategy is used, only stage-2-data must
%be rotated. This, however, would lead to a severe power loss, because it is equivalent to doing a test
%only based on stage 2 data, I think. As a way out, one could, however, rotate stage 1 and stage 2 data separately.

\section{The distribution of Kieser and Friede's $t$-test statistic}\label{sec4}

To investigate the type I error of the $t$-test after a blinded
sample size review, Kieser and Friede \cite{Kieser03} write the
$t$-test statistic as a function of four components $D_1$, $V_1$,
$D_2$ and $V_2^*$ (see page 3575 of \cite{Kieser03}) for which they
derive respective distributions. However, the distribution of
$V_2^*$ given $(D_1,V_1)$ mentioned there is an approximation, not
the exact distribution. Hence, the "actual" type I error rates in
\cite{Kieser03} are also approximate, possibly masking a minor type
I error level inflation.

\medskip
The following uses the notation from \cite{Kieser03}. It shows that
the conditional distribution of $V_2^*|(V_1,D_1)$ is not $\chi^2
(2n_2)$.

Without loss of generality it can be assumed that $\sigma^2=1$. We
have
\[
V_2^*=V_2+\frac{n_1n_2}{n_1+n_2}\left(\left(\overline{X}_{11}-\overline{X}_{21}\right)^2+
\left(\overline{X}_{12}-\overline{X}_{22}\right)^2\right)
\]
$V_2|(V_1,D_1)\sim \chi^2(2n_2-2)$ is obvious. It is also obvious
that if we condition on $V_1$ only, and suppose that this determines
sample size $n_2$ uniquely, we have
\[
D^*_i:=\sqrt{\frac{n_1n_2}{n_1+n_2}}\left(\overline{X}_{1i}-\overline{X}_{2i}\right)\sim
N(0,1),
\]
such that $D^*_1$ and $D^*_2$ are stochastically independent. Thus,
in this case $D^{*2}_1+D^{*2}_2\sim \chi^2(2)$, so if $n_2$ is a
function of $V_1$, but not $D_1$, the claim $V_2^*|V_1\sim
\chi^2(2n_2)$ holds. This was noted by \cite{Gould92}.

If we condition on both $V_1$ and $D_1$, $V_2$ and
$\left(D^*_1,D^*_2\right)$ are still independent, but $D^*_1$ and
$D^*_2$ are no longer.

By applying a theorem on conditional normal distributions (see e.g.
\cite{Sri02}, page 35) and some well-known results on matrix
decompositions, it can be shown that the true conditional
distribution of $V_2^*$ is a mixture distribution:
\[
 V_2^*|(V_1, D_1) =_d \chi^2_{2n_2-1}+ z_2^2,
\]
where "$=_d"$ denotes "equal in distribution" and $z_2^2$ has the
"rescaled" non-central $\chi^2$-distribution
\[
z_2^2\sim
\frac{n_1}{n_1+n_2}\cdot\chi^2\left(1;\frac{n_2}{n_1}\left(D_1-\sqrt{\frac{n_1}{2}}\Delta\right)^2\right).
\]

The assumption $V_2^*|(V_1,D_1)\sim \chi^{2}(2n_2)$ will often very
closely approximate this real distribution.

\section{Sample size reviews when testing for non-inferiority} \label{sec5}

The preceding sections have dealt with the superiority test $H_0:
\mu=0$. While type I error violations in this context are extremely
small, it was noted by \cite{Friede03} that more serious violations
arise in the case of non-inferiority and equivalence testing and
that these are persistent with larger sample sizes. This section
gives an intuitive explanation for this.

Assume that in the two-sample case, it is intended to test the
non-inferiority hypothesis $H_0:\mu_1-\mu_2\leq \delta$ on data
$x_{ijk}\sim N\left(\mu_j,\sigma^2\right)$ where $i=1,2$ indexes
stage, $j=1,2$ treatment group, $k=1,\ldots,n_i$ sample unit and
$\delta$ is a fixed non-inferiority margin. Sample size reassessment
after stage 1 determines the stage-2-sample size via
\begin{equation}\label{f50}
n_2=4\cdot\frac{\left(u_{1-\alpha}+u_{1-\beta}\right)^2}{\left(\theta-\delta\right)^2}\cdot
\tilde{\sigma}^2
\end{equation}
(where $u_\alpha$ is the
$\alpha$-quantile of $N(0,1)$, $\beta$ is the desired power of the
ordinary two-sample $t$-test and $\theta$ is the assumed true effect
difference between the treatments) as a function of the "total
variance"
\[
\tilde{\sigma}^2=\frac{1}{2n_1-1}\sum_{j=1}^{2}\sum_{k=1}^{n_1}\left(x_{1jk}-\bar{x}_1\right)^2
\]
with $\bar{x}_1=\frac{\sum_{j=1}^{2}\sum_{k=1}^{n_1}
x_{1jk}}{2n_1}$. This, however, does not correspond to a "blinded"
sample size review of the corresponding superiority test. To see
this, notice that the described test can also be represented as a
test of $H_0:\mu_1^*-\mu_2\leq 0$ on the "shifted" data
\begin{equation}\label{f51}
x_{ijk}^*=\left\{\begin{array}{l} x_{ijk}-\delta \mbox{ if } j=1,\\
x_{ijk} \mbox{ if } j=2,
\end{array} \right.
\end{equation}

A blinded sample size review of $\left(x_{ijk}^*\right)$ would also
use (\ref{f50}), but with
\[
\hat{\sigma}^2=\frac{1}{2n_1-1}\sum_{j=1}^{2}\sum_{k=1}^{n_1}\left(x^*_{1jk}-\bar{x}^*_1\right)^2
\]
instead of $\tilde{\sigma}^2$. It is easy to see that
\[
\tilde{\sigma}^2=\hat{\sigma}^2-\frac{n_1}{2(2n_1-1)}\delta^2+\frac{n_1}{2n_1-1}\delta(\bar{x}_{11}-\bar{x}_{12}).
\]
This formula contains the quantity
$\frac{n_1}{2n_1-1}\delta(\bar{x}_{11}-\bar{x}_{12})$ which links
the realized difference in means $\bar{x}_{11}-\bar{x}_{12}$ with
the true difference $\delta$ of means under $H_0$. If, for example,
$\delta<0$, then $n_2$ decreases with increasing realized values of
$\bar{x}_{11}-\bar{x}_{12}$. Relative to the blinded superiority
sample size review, this means that fewer additional sample elements
are taken when stage-1-evidence is in favor of the alternative and
vice versa. Obviously, this must be associated with an increase of
type I error under $H_0$. Conversely, the test gets conservative
when $\delta>0$. These tendencies were also noticed by
\cite{Friede03} in simulations.

The "blinded" non-inferiority test is thus equivalent to an
"unblinded" superiority test and hence subject to type I error
biases that afflict an unmodified $t$-test applied after the sample
size was modified using the observed difference in means. To be
sure, the user of the blinded non-inferiority re-estimation does not
get to see the realized value of $\bar{x}_{11}-\bar{x}_{12}$, but
nevertheless it has the described impact on the modified sample size
$n_2$.

%Hence, when we represent the "blinded" non-inferiority investigation
%in its equivalent superiority form.

\section{Discussion}

This paper investigates a number of situations with normally
distributed observations where blinded sample size review according
to Kieser and Friede does not control the type I error rate. In
superiority testing, the corresponding inflations are extremely
small and occur with sample sizes that will rarely be of practical
relevance. The method can thus safely be used in practice.

As an alternative for which type I error control can be proved, it
is also possible to combine the $t$-test statistics of the two
stages directly using data-dependent weights. Regarding the outcome
in practical applications, these two methods are virtually
indistinguishable. In contrast, $p$-value-combination and related
methods suffer from some power loss due to the fact that they have
to work with a predetermined "intended" stage-2-sample size and lose
power if one deviates from this intention in the sample size review.

%True, but I think this does not belong here.
%I don't see how one could possibly calculate of a putative "worst case" in this context without getting very conservative.
%Note also that adjustment for a putative "case of worst type I error
%inflation" is not an option. As shown by \cite{Proschan95}, this
%might result in extremely conservative adjustments.

Non-inferiority testing is subject to much more severe type I error
violations. This is due to its equivalence with unblinded
superiority testing. As a consequence, blinded SSR is not an
acceptable method in confirmatory clinical trials.

\section{Technical Appendix}

This appendix shows that $V_2^*|(V_1, D_1)$ has the distribution
given in section \ref{sec4}. By applying the usual theorems on
conditional normal distributions (see e.g. \cite{Sri02}, page 35),
we obtain the bivariate distribution
\begin{eqnarray} \label{f3}
&&\mathbf{D}^*:={D^*_1 \choose D^*_2}\left|(V_1,D_1)\right. ={D^*_1
\choose
D^*_2}\left|(n_2,D_1)\right.\sim\\
&&N_2\left({\sqrt{\frac{n_2}{2(n_1+n_2)}}
\left(D_1-\sqrt{\frac{n_1}{2}}\Delta\right) \choose
{-\sqrt{\frac{n_2}{2(n_1+n_2)}}}
\left(D_1-\sqrt{\frac{n_1}{2}}\Delta\right)}; \left(
\begin{array}{cc}
\frac{2n_1+n_2}{2(n_1+n_2)} & \frac{n_2}{2(n_1+n_2)} \\
\frac{n_2}{2(n_1+n_2)} & \frac{2n_1+n_2}{2(n_1+n_2)}
\end{array}
\right)\right)\nonumber
\end{eqnarray}

To derive the distribution of
$\mathbf{D}^{*'}\mathbf{D}^*=D^{*2}_1+D^{*2}_2$, we can make use of
the following well-known general result:

Suppose $\mathbf{x}\sim N_p(\bmu,\mathbf{V})$ and let
$\mathbf{V}^{\frac{1}{2}}$ be a root of $\mathbf{V}$ (i.e. a matrix
that fulfills
$\mathbf{V}^{\frac{1}{2}}\mathbf{V}^{\frac{1}{2}}=\mathbf{V}$). Then
$\mathbf{x}=_d\mathbf{V}^{\frac{1}{2}}\mathbf{y}$ where
$\mathbf{y}\sim N_p(\mathbf{V}^{-\frac{1}{2}}\bmu,\mathbf{I}_p)$.

Furthermore assume that $\mathbf{A}$ is a positive semidefinite
symmetric $p\times p$-matrix. Then
\begin{equation}\label{f10}
 \mathbf{x}'\mathbf{A}\mathbf{x}=_d
\mathbf{y}'\mathbf{V}^{\frac{1}{2}}\mathbf{A}\mathbf{V}^{\frac{1}{2}}\mathbf{y}.
\end{equation}
$\mathbf{V}^{\frac{1}{2}}\mathbf{A}\mathbf{V}^{\frac{1}{2}}$ can
also be written as an eigenvalue decomposition
$\mathbf{C}'\mathbf{\Lambda}\mathbf{C}$, where
$\mathbf{\Lambda}=\left(\lambda_i\right)_{i=1,\ldots,p}$ is the
diagonal matrix of eigenvalues and $\mathbf{C}$ is the matrix of the
corresponding eigenvectors. Inserting this into (\ref{f10}), we
obtain
\[
\mathbf{x}'\mathbf{A}\mathbf{x}=_d \sum_{i} \lambda_i z_i^2
\]
with $\mathbf{z}=(z_1,\ldots, z_p)'\sim
N\left(\mathbf{C}'\mathbf{V}^{-\frac{1}{2}}\bmu,
\mathbf{I}_p\right)$.

Using this general result in the particular case by setting $\bmu$
and $\mathbf{V}$ to the mean and covariance matrix in ($\ref{f3}$)
and $\mathbf{A}= \mathbf{I}_2$, it is easy to see that $\mathbf{V}$
has eigenvalues 1 and $\frac{n_1}{n_1+n_2}$ and eigenvectors
$\frac{1}{\sqrt{2}}\cdot (1,1)'$ and $\frac{1}{\sqrt{2}}\cdot
(1,-1)'$. Consequently, conditional on $(V_1,D_1)$, we obtain:
\[
D^{*2}_1+D^{*2}_2=_d z^{2}_1+z^{2}_2
\]
where $z_1\sim N(0,1)$ and $z_2\sim
N(\sqrt{\frac{n_2}{n_1+n_2}}\cdot
\left(D_1-\sqrt{\frac{n_1}{2}}\Delta\right),\frac{n_1}{n_1+n_2})$.
Hence, $V_2^*|(V_1, D_1) =_d \chi^2_{2n_2-1}+ z_2^2$, where $z_2^2$
has the "rescaled" non-central $\chi^2$-distribution
\[
z_2^2\sim
\frac{n_1}{n_1+n_2}\cdot\chi^2\left(1;\frac{n_2}{n_1}\left(D_1-\sqrt{\frac{n_1}{2}}\Delta\right)^2\right).
\]

We note in passing that if $\mathbf{D}^{*'}\mathbf{D}^*$ were
$\chi^2(2)$-distributed, it would have
$E(\mathbf{D}^{*'}\mathbf{D}^*)=2$. The true conditional expected
value given $(V_1, D_1)$ can be obtained from
\begin{eqnarray*}
E(\mathbf{D}^{*'}\mathbf{D}^*)=tr(E(\mathbf{D}^*\mathbf{D}^{*'}))=tr(\mathbf{\Sigma}_{\mathbf{D}^*}+\bmu_{\mathbf{D}^*}\bmu_{\mathbf{D}^*}')=\\
1+\frac{n_1}{n_1+n_2}+\frac{n_2}{n_1+n_2}\left(D_1-\sqrt{\frac{n_1}{2}}\Delta\right)^2.
\end{eqnarray*}

Of course, this is not equal to $2$ in general. However,
$E(\left(D_1-\sqrt{\frac{n_1}{2}}\Delta\right)^2)=1$ holds, since
$D_1 \sim N\left(\sqrt{\frac{n_1}{2}}\Delta,1\right)$. If we then
ignore that $n_2$ is a random variable as well, we obtain the
approximate unconditional expected value
$\left(1+\frac{n_1}{n_1+n_2}+\frac{n_2}{n_1+n_2}\right)=2$.

\end{document}